\documentclass[a4paper,11pt]{article}
\usepackage{pdflscape}
\usepackage{amsmath}
\usepackage{amssymb}
\usepackage{dcolumn}
\usepackage{bm}
\usepackage{color}
\usepackage{epsfig}
\usepackage{amsfonts}
\usepackage{graphicx}
\usepackage{subfigure}
\usepackage{dcolumn}
\usepackage{indentfirst}
\usepackage{authblk}
\usepackage{slashed}
\usepackage{appendix}

\usepackage{booktabs}
\AtBeginDocument{
\heavyrulewidth=.08em
\lightrulewidth=.05em
\cmidrulewidth=.03em
\belowrulesep=.65ex
\belowbottomsep=0pt
\aboverulesep=.4ex
\abovetopsep=0pt
\cmidrulesep=\doublerulesep
\cmidrulekern=.5em
\defaultaddspace=.5em
}

 \usepackage{ulem}
\begin{document}


\vspace{80pt}

\centerline{\LARGE Conditions for vacuum instability in holographic theories }

\vspace{10pt}

\centerline{\LARGE with dilaton field}

\vspace{40pt}

\centerline{
Sara Tahery,$^{a}$ 
\footnote{E-mail: s.tahery@impcas.ac.cn}
Sreeraj Nair,$^{b}$ 
\footnote{E-mail: sreeraj@impcas.ac.cn}
and
Pengming Zhang $^{c}$ 
\footnote{E-mail: zhangpm5@mail.sysu.edu.cn}
}
\vspace{30pt}

{\centerline {$^{a,b}${\it Institute of Modern Physics, Chinese Academy of Sciences, Lanzhou,  China}}

\vspace{4pt}

{\centerline {$^{c}${\it School of Physics and Astronomy, Sun Yat-sen University, Zhuhai, China}}

 \vspace{40pt}


\begin{abstract}
We investigate the vacuum instability in the presence of 
dilaton field in  a holographic
set up. Although the dilaton is a bulk field, it leads to  the vacuum 
instability on the boundary.
 We 
show that the whole
process crucially depends on the probe brane position  and  as well on 
the radial coordinate.
So that the effects of dilaton scale parameter in different regions of 
the bulk or for different
probe brane positions are different. We also observe that in our study 
the temperature can
strengthen the effect of scale parameter in reducing the potential 
barrier. Finally, we show
that this  Schwinger-like effect,  although is interesting by itself,  
does not produce a considerable  pair production rate.
\end{abstract}

\newpage

\tableofcontents

\newpage

\section{Introduction}
Pair production in presence of an external electric field is known as the Schwinger effect in non-perturbative quantum electrodynamics (QED) \cite{SCH}. Due to this phenomenon when the external field is strong enough, the virtual electron-positron pair become real particles. In other words, vacuum is destroyed in presence of such a field.\\
Although this context had been considered in QED first, it is not restricted to it any more. It has been extended to quantum chromodynamics (QCD) and even higher dimensional objects like strings and branes \cite{uah}. As an interesting example, it has been considered in a gravitational wave background while  the electric field was replaced by the the gravitational wave background and the electron/positron field quanta was
replaced by massless scalar photons \cite{wave}.
In other words nowadays the Schwinger effect is not restricted to QED, external electric field and electron-positron pair production, but any kind of vacuum decay due to pair-production in presence of any external field stands for the Schwinger effect.

Potential analysis plays an important role in  vacuum stability. The main set up of our current work is based on considering vacuum decay process by diminishing potential barrier. Before going to the main subject we mention Schwinger effect which gives us the strategy of our computation, but we will discuss the important differences between this effect (in its known form) and the current work in upcoming calculations as well.
 
In the context of QED, the potential analysis
estimated by the static potential includes the Coulomb interaction between the particles in addition to an energy $Ex$, where $x$ is a separating distance of virtual pair and $E$ is an external electric field  \cite{hds}. Generally  the total potential is calculated by the Lagrangian integration over the internal distance of the  pair, in addition to a term coming from the external field energy. This is the strategy: the internal energy of the virtual pair leads to  a potential barrier. When the virtual pair get a greater energy
than the rest energy from an external field, they become real. So, for the creation of a real pair which corresponds to the vacuum decay, the external field should reach to a critical value, where the vacuum becomes totally unstable.

 In QED, increasing electric field can destroy potential barrier and finally vacuum decays. Accordingly, one can expect that in any  kind of Schwinger effect,  increasing external field results in destroying potential barrier.
  When the external field is small, the potential barrier is present and the pair production is a tunneling process. The potential barrier diminishes as the external field increases. At a critical value of the field, the potential barrier vanishes completely and the pair production is catastrophic \cite{uah}.
From string theory point of view, this critical value is regarded as string behaviour in ultraviolet (UV) completion of the string \cite{qst,pcr}. 

One strong tool to investigate Schwinger effect in string context or higher dimensional objects is AdS/CFT which is a  correspondence describing a  relation between a d-dimensional conformal field theory (CFT) and a $(d + 1)$ dimensional string theory in anti-de Sitter (AdS) space \cite{mal}. This is a powerful mathematical tool to investigate about strongly correlated systems \cite{wit, gau, nfi}. Extra
dimension in the AdS side leads to using the energy scale of the CFT side on
the boundary.
Although QCD is not a CFT exactly, but in recent decades AdS/QCD has been considered as an useful approach to study an analytic semi-classical model for strongly
coupled QCD. It has scale invariance, dimensional counting at short distances and color confinement at large distances. This theory describes the phenomenology of hadronic properties and demonstrates their ability to incorporate such essential properties of QCD like confinement and chiral symmetry breaking. From the AdS/CFT point of view the $AdS_5$ plays an important role in describing QCD phenomenon so it is called AdS/QCD \cite{lic,adsqcd,corr}.

 Many works have been done about Schwinger effect in a holographic setup related to quark-antiquark pair production as follows, the creation rate of the quark pair in $N=4$ SYM theory was obtained in \cite{stp} and based on that, the holographic  Schwinger effect was calculated in various systems \cite{uah,hds, hawk,rhw,coce,secp,ppr,hdse,npp,hsed,nre,scen,udit}. Also the vacuum decay rate is regarded as the creation rate of the quark-antiquark in $N=2$ supersymmetric QCD (SQCD) \cite{vie}. In the Ref.\cite{pah} electrostatic potentials in the holographic Schwinger effect has been analyzed for the finite-temperature and temperature-dependent critical-field cases to find  agreement with the full form Dirac-Born-Infeld (DBI) result. In  Ref.  \cite{hse} tunneling pair creation of W-Bosons by an external electric field on the Coulomb branch of $N=4$ super symmetric Yang-Mills theory has been studied and found that the pair creation formula has an upper critical electric field beyond which the process is no longer exponentially suppressed. 
 
 Light Front Holographic QCD \cite{lfh,lfhq}  is a model theory, which tries to explain non-perturbative features of quantum field theory for strong interactions, QCD.
 In order to get some insight into the structure of the most interesting phenomena, one has to make specific models and approximations. An important approach is the semi-classical approximation of a quantum field theory.
The basis of light front holographic QCD is the ``holographic principle" which states that certain aspects of a quantum field theory in four space-time dimensions can be obtained as limiting values of a five dimensional theory as it is mentioned before. In Light Front Holographic QCD (LFHQCD) one chooses a bottom-up approach, that is one modifies the five dimensional classical theory in such a way as to obtain from this modified theory and the holographic principle realistic features of hadron physics \cite{1801}.
In LFHQCD, the action is an invariant action, modified by a dilaton term $e^{\varphi(z)}$ as
\begin{align}
S_{\rm eff}=&\int d^dx dz \sqrt{g} e^{\varphi(z)} g^{N_1 N'_1}...g^{N_j N'_j}(g^{MM'}D_M \Phi^{*}_{N_1...N_j} D_{M'} \Phi_{N'_1...N'_j}\nonumber\\
&-\mu^{2}_{eff}(z) \Phi^{*}_{N_1...N_j} \Phi_{N'_1...N'_j}),
\end{align}
according to the dictionary between the AdS result and the LFH the potential is related to the dilaton field in the effective $AdS_5$ action.
The corresponding metric with the mentioned action is an asymptotic $AdS_5$
metric modified by a dilaton field $\varphi(z)$. It is only a function of the holographic variable $z$ which vanishes in the conformal limit $z \rightarrow 0$. In $AdS_5$, this unique $z$-dependence of the dilaton field allows
the description of the bound-state dynamics in terms of a one dimensional wave equation. It also enables one to establish a map to the semi-classical one-dimensional approximation to light-front QCD given by the frame-independent light-front Schr\"{o}dinger equation. It has been found that the dilaton profile has the specific form: $\varphi(z)=-\lambda z^2$ \cite{1801} which leads to linear Regge trajectories and avoids the ambiguities in
the choice of boundary conditions at the infrared wall \cite{Brodsky:2014yha}. The spectrum can only be described by choosing $\lambda>0$. Thus, in this work we consider the dilaton profile as $\varphi(z)=-\lambda z^2$ with positive $\lambda$.

In Ref. \cite{stsch} one of the authors of this work considered vacuum instability in a deformed AdS in presence of an electric field. In current work, our motivation is to consider vacuum instability by this holographic model and without any external electric field. This holographic model is important from two different points of view. First, it stands for light front holographic approach which has been mentioned before. Second, it can be considered as a deformed AdS where one deforms the AdS by second correction of radial coordinate \cite{cor}, to discuss on some asymptotically AdS behaviour of the theory. In the next section we will represent such a metric with an estimate of the quadratic correction of radial coordinate based on gauge/string duality. In references \cite{cor, hqp} it is written that in QCD analysis of the two current correlator the first coefficient can be calculated perturbatively, while the second coefficient (quadratic correction) is not easy to be found. Then it should be estimated based on some data. Correspondingly in gauge/gravity duality, in models with the slightly deformed $AdS_5$ metric, one can finds some results fit better to QCD ramains to be seen. Here we only addressed the main motivation and avoided to open  many details of discussion, so interested reader can refer to the main references. We will focus on the first point of view keeping in our mind that our results will cover deformed AdS/QCD too \cite{hqp}. So, starting by the soft-wall LFH metric, we are interested in studying vacuum decay process. 

The process starts from ``turning on the $\lambda$'' means to consider non zero value of this scale parameter. As $\lambda$ increases, we expect that potential barrier diminishes. Therefore one interprets that the potential barrier is supposed to be vanished by ``large enough value of $\lambda$''. Although Schwinger effect has been considered by external electric field and magnetic field before, the most important difference of current work is that the vacuum decay initiates from inside the metric.  Briefly, $\lambda$ is responsible for vacuum decay and pair production, thus it has the main role in this process. This is a goal to see the effects of space-time specifications during vacuum decay.

With all above explanations we represent this paper as follows, in section 2 we  consider vacuum decay by dilaton field at zero temperature. Proceeding by finite temperature we follow the study in section 3. In sections 4 and 5 the pair production rate for both zero and thermal cases are discussed. Section 6 is the numerical strategy and our conclusion and results will be represented in section 7.

\section{Potential analysis at zero temperature}

Considering LFH metric at zero temperature, we analyze potential initiated by the dilaton field.
According to the holographic set up in \cite{pah} we will derive the total potential from the action. The LFH metric is written as, 
\begin{equation} \label{metric zero}
ds^2=\frac{R^2}{z^2} h(z) (-dt^2+\Sigma_{i=0}^3 dx_i ^2+dz^2)+R^2 d\Omega_{5}^2,\quad \quad h(z)=e^{-\lambda z^2},
\end{equation}
where $R$ is the radius of space which is related to the slope parameter and coupling via, $R^2=\alpha' \sqrt{\lambda}$, with $\alpha'=l_s^2$ where $l_s$ is the string scale. Moreover $d\Omega_5^2$ is the metric of a five-dimensional sphere.

The potential of the produced pair particle is obtained using the expectation value of the Wilson loop.
The loop  corresponds to a trajectory of test particles with infinite heavy mass, and the expectation value  corresponds to the area of a string world-sheet attached to the Wilson loop \cite{mas,wil}.
Thus, in order to study by AdS/CFT the area of rectangular Wilson loop on the probe D3-brane evaluates classical action of a string attached to the probe D3-brane \cite{hse}. The Nambu-Goto
string action is given by,
\begin{eqnarray} \label{eq:NG}
S&=&T_F \int {d\tau d \sigma \mathcal{L}}\nonumber\\
&=&T_F  \int {d\tau d \sigma \sqrt{det G_{ab}}},
\end{eqnarray}
where
\begin{equation}\label{G}
G_{ab}\equiv \frac{\partial x^\mu}{\partial \sigma^{a}} \frac{\partial x^\nu}{\partial \sigma^{b}} g_{\mu\nu},
\end{equation}
is the induced metric and $\sigma^{a}=(\tau, \sigma)$ are world-sheet coordinates  and $T_F=\frac{1}{2\pi \alpha'}$ is the string tension. From the relation (\ref{metric zero}) we have,
\begin{equation} \label{eq:eta zero-c}
g_{ab}=\mathrm{diag}\left(-\frac{R^2}{z^2}e^{-\lambda z^2}, \frac{R^2}{z^2}e^{-\lambda z^2}\right).
\end{equation}
It is useful to choose the static gauge, $x^{0}=\tau$, and $x^{1}=\sigma$. So, the radial direction $z(\sigma)$ depends only on $\sigma$ in classical solution.
Therefore, the Lagrangian is,
\begin{equation} \label{eq:L zero}
\mathcal{L}=\frac{R^2}{z^2} e^{-\lambda z^2} \sqrt{1+\left(\frac{dz}{d\sigma}\right)^2}.
\end{equation}
From the equation of motion, one can find,
\begin{equation} \label{eq: eqmo}
\frac{\partial \mathcal{L}}{\partial (\partial _{\sigma} z)} \partial_{\sigma} z -\mathcal{L}=C_{1},
\end{equation}
where $C_{1}$ is an arbitrary constant, and this yields to the following relation,
\begin{equation} \label{eq:cte zero}
\frac{R^2}{z^2}\frac{ e^{-\lambda z^2}}{\sqrt{1+\left(\frac{dz}{d\sigma}\right)^2}}=C_{2},
\end{equation}
again, $C_{2}$ is an arbitrary constant. The important boundary condition at $\sigma=0$, imposes,
\begin{equation} \label{eq:bouncon zero}
\frac{dz}{d\sigma}=0, \quad \quad z=z_{\ast},
\end{equation}
where $z_{\ast}$ is the turning point, which means the deepest position of the string in the bulk. Therefore, we yield to the following differential equation,
\begin{equation} \label{eq:difeq zero}
\frac{dz}{d\sigma}=\sqrt{\frac{z_\ast ^4}{z^4} e^{-2\lambda (z^2-z_\ast ^2)}-1} .
\end{equation}
Thus, the separation length of the test particles on the probe brane is,
\begin{equation}\label{eq:x zero}
x=\int_{z_0}^{z_{\ast }}\frac{dz}{\sqrt{\frac{z_{\ast }^4}{z^4} e^{-2\lambda (z^2-z_\ast ^2)}-1}},
\end{equation}
where $z_0$ is the probe brane position. We will use this length in considering behaviour of potential later. From the Lagrangian (\ref{eq:L zero}) the potential of the produced pair particle is given by,
\begin{eqnarray}\label{eq:vcp zero}
V&=&2T_F\int_{0}^{\frac{x}{2}} dx \mathcal{L}\nonumber\\
&=&2T_F  R^2 z_{\ast }^2 \int_{z_0}^{z_{\ast }} \frac{1}{z^4}\frac{e^{-\lambda (2z^2-z_\ast ^2)}}{\sqrt{ \frac{z_{\ast }^4}{z^4} e^{-2\lambda (z^2-z_\ast ^2)}-1}} dz.
\end{eqnarray}
Before studying this potential, it is useful to introduce another quantity as critical dilaton field. 
The critical value of the dilaton field corresponds to the value  
at which the potential barrier is destroyed and pair production begins. 
According to the metric and by considering the fact that the critical field is interpreted as string tension $\sigma_{\mathrm{string}}$ in string theory side \cite{hash}, 
 \begin{equation}\label{eq:cr}
\sigma_{\mathrm{string}}=T_F\sqrt{-g_{00}g_{11}}\vert_{IR},
\end{equation}
then the result at zero temperature is 
  \begin{equation}\label{eq:cr2}
\sigma_{\mathrm{string}}=T_F \frac{R^2}{z_0^2} e^{-\lambda z_0^2}
\end{equation}
which corresponds to,
\begin{equation}
h(z)_{\mathrm{cr}}=T_F \frac{R^2}{z_0^2} e^{-\lambda z_0^2}.
\end{equation}
We define a dimensionless value as the ratio of the field to its critical value as,
\begin{equation}
 \alpha=\frac{h(z)}{h(z)_{cr}}=\frac{e^{-\lambda z^2}}{T_F \frac{R^2}{z_0^2} e^{-\lambda z_0^2}}=\frac{z_0^2}{T_F R^2} e^{-\lambda (z^2-z_0^2)},
\label{alpha}
\end{equation}

Based on previous works on Schwinger effect,
we expect that when $\alpha$ is unity the pair production process gets started \cite{ppr,pah}. We will see that this is not a sufficient condition for pair production in this work.

\begin{figure}[h!]
\begin{center}$
\begin{array}{cccc}
\includegraphics[width=100 mm]{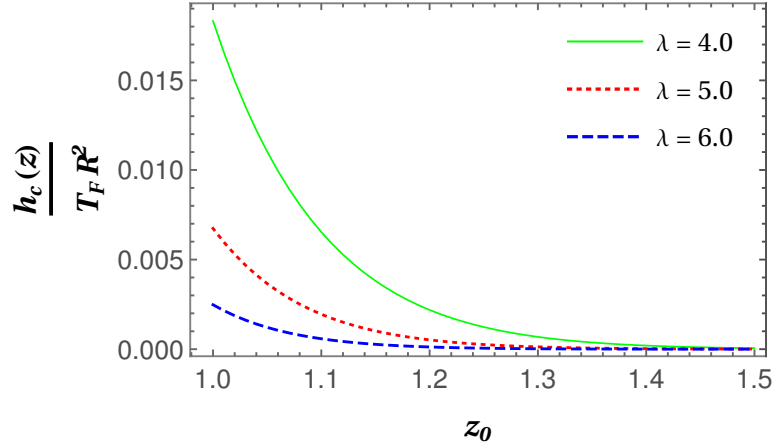}
\end{array}$
\end{center}
\caption{Considering critical field against prob-brane position, at zero temperature and for different values of scale parameter.}
\label{fig:critical_vs_z0_different_lambda}
\end{figure}

It is worth mentioning that, the space-time metric parameters and the brane configuration all affect the Schwinger effect considered here.
Fig. \ref{fig:critical_vs_z0_different_lambda} considers the critical value of the field as a function of $z_0$. Clearly, with increasing $z_0$ the critical value of the field decreases. So when the probe brane is near boundary ($z=0$), larger value of critical field is obtained for pair production. In addition, greater $\lambda$ corresponds to smaller value of critical field at the same $z_0$. This behaviour has an exception in the region far from the boundary where different plots with different $\lambda$ are coincident. It means that when the probe brane is far from boundary, scale parameter has no significant effect on critical field. 
 
\begin{figure}[h!]
\begin{minipage}[c]{1\textwidth}
\begin{center}
\tiny{(a)}\includegraphics[width=7cm,height=6cm,clip]{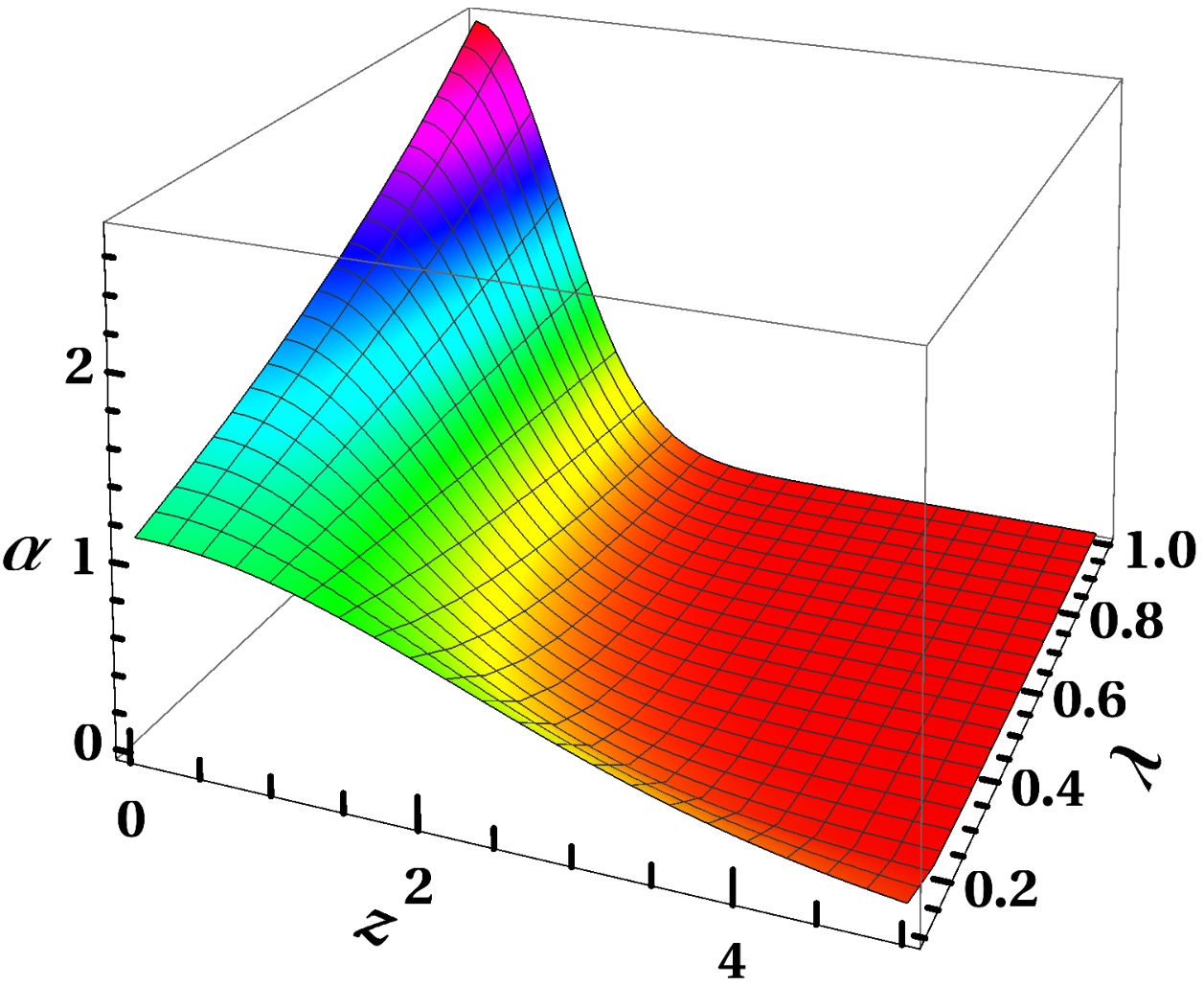}
\tiny{(b)}\includegraphics[width=7cm,height=6cm,clip]{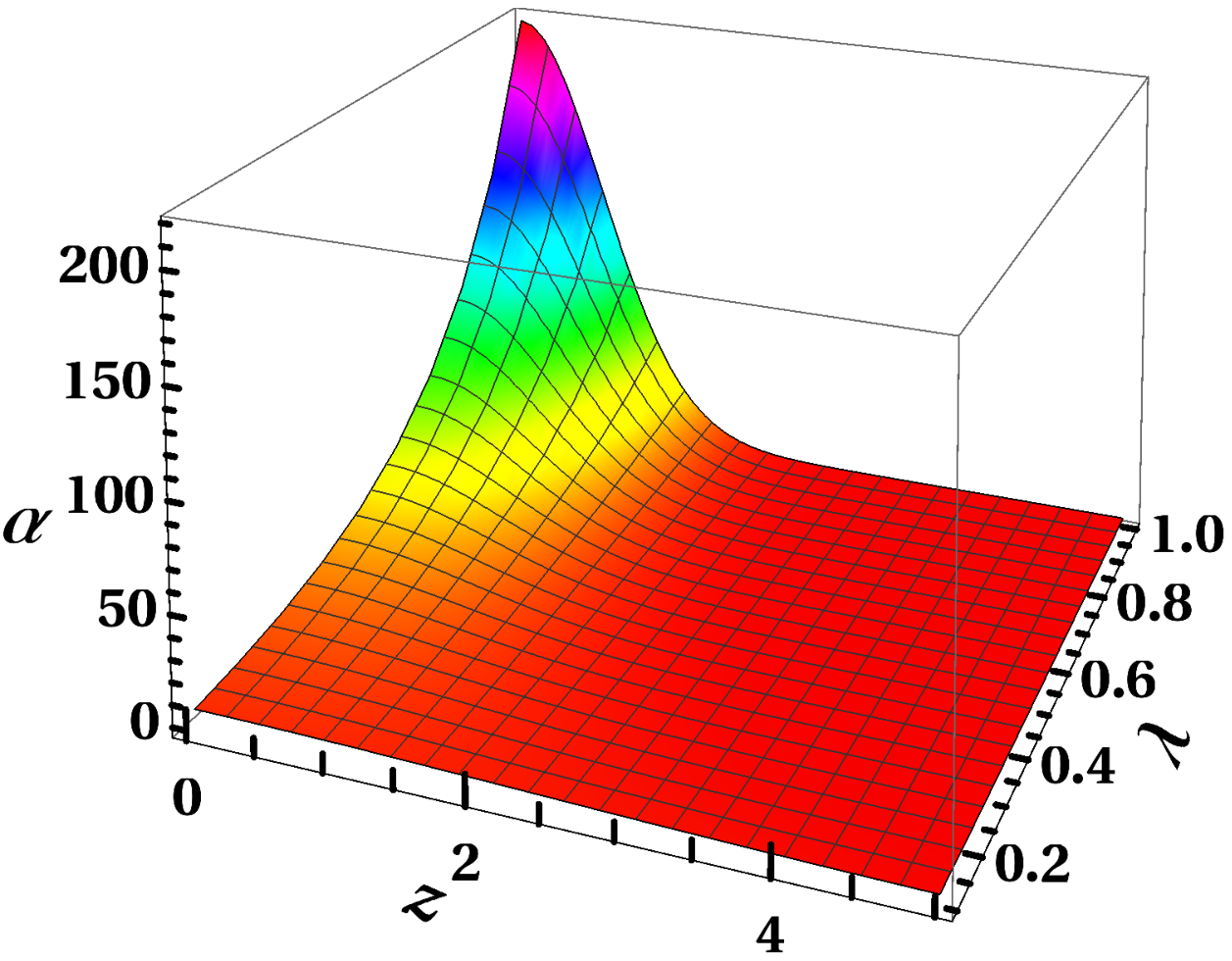}
\end{center}
\end{minipage}
\caption{Considering pair production parameters versus each other in 3D plot, at zero temperature and fixed value of probe brane position at (a) $z_0=1$, (b) $z_0=2$.} 
\label{fig:3dplot_alpha_vs_lambda_z_fixed_z0}
\end{figure}

In Fig. \ref{fig:3dplot_alpha_vs_lambda_z_fixed_z0},  based on the relation between $\alpha$ and $\lambda$, behaviour of $\alpha$  has been considered with respect to the scale parameter $\lambda$, for different values of $z_0$. By choosing fixed probe brane position, one can consider effects of scale parameter and position of probe brane on $\alpha$. Obviously, when the probe brane is  in near boundary region $\alpha=1$  is obtainable in a limited region near it and for a wide range of $\lambda$ values as we can see in  plot (a) . 
By increasing $z_0$, in plot (b)  the condition $\alpha\geq 1$ is satisfied almost along $z$ coordinate. With increasing $z_0$ large values of scale parameter leads to large values of $\alpha$.




\begin{figure}[h!]
\centering
\includegraphics[width=100 mm,clip]{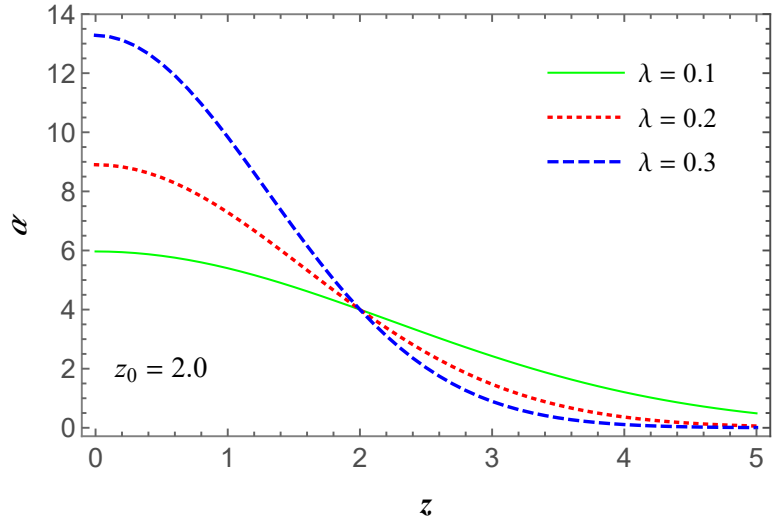}
\caption{Considering behaviour of $\alpha$ along the axis $z$ at zero temperature with different values of scale parameter.}
\label{fig:Alpha_vs_z}
\end{figure}

In Fig. \ref{fig:Alpha_vs_z} , we show the two dimenensional (2D) cross-section of Fig. \ref{fig:3dplot_alpha_vs_lambda_z_fixed_z0} for fixed probe-brane position $z_0 =2$. We observe a critical point at $z=z_0$ where $\alpha$ is independent of $\lambda$.
After the critical point,  $\alpha$ decreases with  increasing $\lambda$.
 In other words not only scale parameter affect pair production but also the probe brane position is important. In addition, the behaviour depends on the distance along $z$ axis. 

\begin{figure}[h!]
\begin{center}$
\begin{array}{cccc}
\includegraphics[width=100 mm]{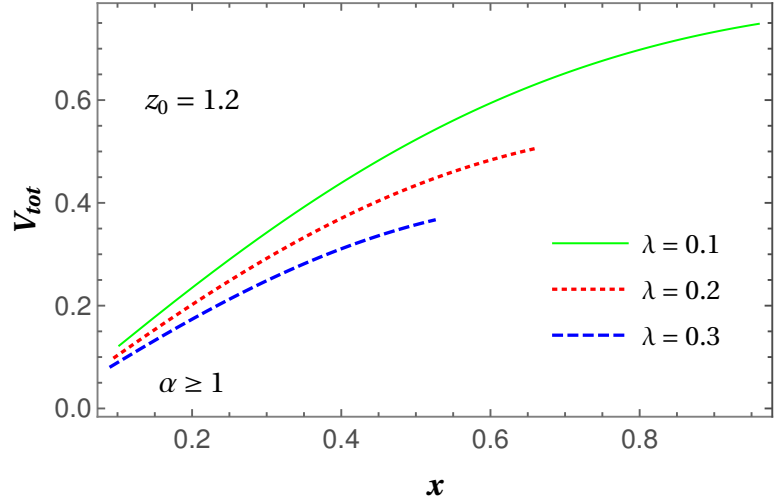}
\end{array}$
\end{center}
\caption{Considering behaviour of potential during pair production at zero temperature.}
\label{fig:x_vs_vtot_z0_1.2_alphage1}
\end{figure}

\begin{table}[ht]
\centering
\begin{tabular}[t]{lccc}
\toprule
~$\lambda$~~ & $x$~~ & $V$~~ \\
\midrule
~0.1 & 0.96   &  0.75\\
~0.5 & 0.38   &  0.21\\
~1.0 & 0.24   &  6.59 $\times$ $10^{-2}$\\
~2.0 & 0.13   &  9.18 $\times$ $10^{-3}$\\
~5.0 & 0.06   &  5.63 $\times$ $10^{-5}$\\
\bottomrule
\end{tabular}
\label{1}
\caption{Maximum values of  $V_{\mathrm{tot}}$ and $x$ for different values of $\lambda$ for $z_0 = 1.2$.}
\end{table}%

Fig. \ref{fig:x_vs_vtot_z0_1.2_alphage1} shows the potential for different $\lambda$ values with a fixed probe brane position. 
We observe that $\alpha \ge 1$ is not sufficient condition for pair production, we need large value of $\lambda$ to overcome the potential barrier.
Table 1 shows the maximum values of  $V_{\mathrm{tot}}$ and $x$ for different values of $\lambda$. One notices that for large $\lambda$, the potential vanishes. So we can conclude that to create a pair the necessary condition is to have large $\lambda$. 

\section{Potential analysis at finite temperature}
In this section potential analysis will be considered at finite temperature. The modified thermal metric is given by,
\begin{equation}\label{eq:metric finite t}
ds^2=\frac{R^2}{z^2} h(z)  (-f(z) dt^2+\Sigma_{i=1}^{3}dx_i^2+\frac{1}{f(z)} dz^2)+R^2 d\Omega_5^2,
\end{equation}
where
\begin{equation}\label{eq:metric function finite t}
f(z)=1-\left(\frac{z}{z_h}\right)^4,\quad\quad h(z)=e^{-\lambda z^2}.
\end{equation}
The horizon is located at $z=z_h$ where $z_0<z_{\ast}<z_h $, and the temperature of the black hole is written as, $T=\frac{1}{\pi z_h}$, so zero temperature limit $z_{h}\rightarrow\infty$ and $f(z)\rightarrow 1$ has been discussed in the previous section.
The Lagrangian is given by,
\begin{equation} \label{eq:L T}
\mathcal{L}=\frac{R^2}{z^2} e^{-\lambda z^2} \sqrt{\left(1-\frac{z^4}{z_h^4}\right)+\left(\frac{dz}{d\sigma}\right)^2}.
\end{equation}
By using the equation of motion, one can find,
\begin{equation} \label{eq:cte t}
\frac{R^2}{z^2}\frac{ e^{-\lambda z^2}(1-\frac{z^4}{z_h^4})}{\sqrt{\left(1-\frac{z^4}{z_h^4}\right)+\left(\frac{dz}{d\sigma}\right)^2}}=C,
\end{equation}
which yields to the following differential equation,
\begin{equation} \label{eq:difeq t}
\frac{dz}{d\sigma}=\sqrt{\frac{z_\ast ^4}{z^4} e^{-2\lambda(z^2-z_\ast ^2)}\frac{(1-\frac{z^4}{z_h^4})^2}{(1-\frac{z_{\ast}^4}{z_h^4})}-\left(1-\frac{z^4}{z_h^4}\right)}.
\end{equation}
So, the internal separation length of the pair particles is obtained as,
\begin{equation}
x=\int_{z_0}^{z_{\ast}}\frac{dz}{\sqrt{\frac{z_\ast ^4}{z^4} e^{-2\lambda(z^2-z_\ast ^2)}\frac{\left(1-\frac{z^4}{z_h^4}\right)^2}{\left(1-\frac{z_{\ast}^4}{z_h^4}\right)}-\left(1-\frac{z^4}{z_h^4}\right)}},
\end{equation}

and the total potential is found as,
\begin{eqnarray}\label{eq:vcp zero}
V&=&2T_F\int_{0}^{\frac{x}{2}} dx \mathcal{L}\nonumber\\
&=&2T_F  R^2 z_{\ast }^2 \int_{z_0}^{z_{\ast }} \frac{1}{z^4}\frac{e^{-\lambda (2z^2-z_\ast ^2)}}{\sqrt{ \frac{z_{\ast }^4}{z^4} e^{-2\lambda (z^2-z_\ast ^2)}-\frac{\left(1-\frac{z_{\ast}^4}{z_h^4}\right)}{\left(1-\frac{z^4}{z_h^4}\right)}}} dz.
\end{eqnarray}
Similar to last section, there is a critical value of the field, in which the pair production process starts.  The thermal metric  results in \cite{hash}, 
\begin{equation}\label{eq:alpha}
h(z)_{cr}=T_F \frac{R^2}{z_0^2} e^{-\lambda z_0^2}\sqrt{1-b^4},
\end{equation}
where $b=\frac{z_0}{z_h}$.
Using (\ref{eq:alpha}), one can derive the $\alpha$ in thermal case as,
\begin{equation}
 \alpha=\frac{h(z)}{h(z)_{cr}}=\frac{z_0^2}{T_F R^2} \frac{e^{-\lambda (z^2-z_0^2)}}{\sqrt{1-b^4}}.
\end{equation}

According to this ratio we proceed by considering it in different temperatures and at fixed scale parameter in Fig.~\ref{fig:Alpha_vs_zfinT} (a). At the boundary, $\alpha$  has it's maximum value. Moving along radial coordinate, $\alpha$ decreases significantly. This manner is obvious in thermal case, similar to what we have considered in Fig.  \ref{fig:Alpha_vs_z} at zero temperature.
We observe that $\alpha$ increases with increasing temperature but far from boundary region $\alpha$ becomes independent of temperature.
The behaviour of $\alpha$ at fixed temperature but with different scale parameter is shown in Fig. \ref{fig:Alpha_vs_zfinT} (b)
where we find the effects of $\lambda$ on $\alpha$ according to the region. 
We observe a critical point at $z=z_0$ similar to the zero temperature case at which $\alpha$ becomes independent of $\lambda$. 
The greater $\lambda$ leads to  larger $\alpha$ from boundary up to $z=z_0$ in the bulk then this behaviour changes clearly in reverse. It can be interpreted as the effect of scale parameter near the boundary is completely different with near horizon region.

\begin{figure}[h!]
\begin{minipage}[c]{1\textwidth}
\tiny{(a)}\includegraphics[width=7cm,height=6cm,clip]{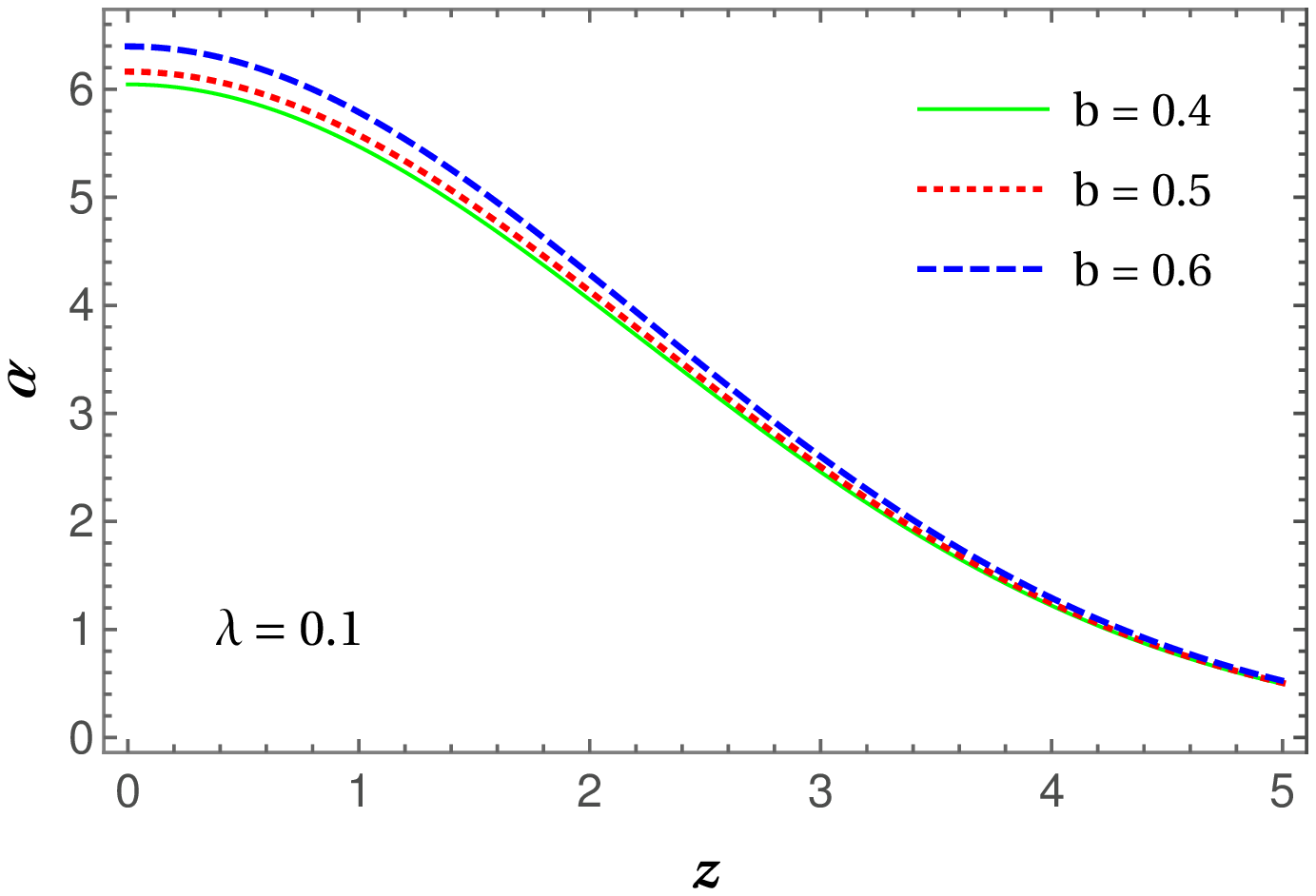}
\hspace{0.1cm}
\tiny{(b)}\includegraphics[width=7cm,height=6cm,clip]{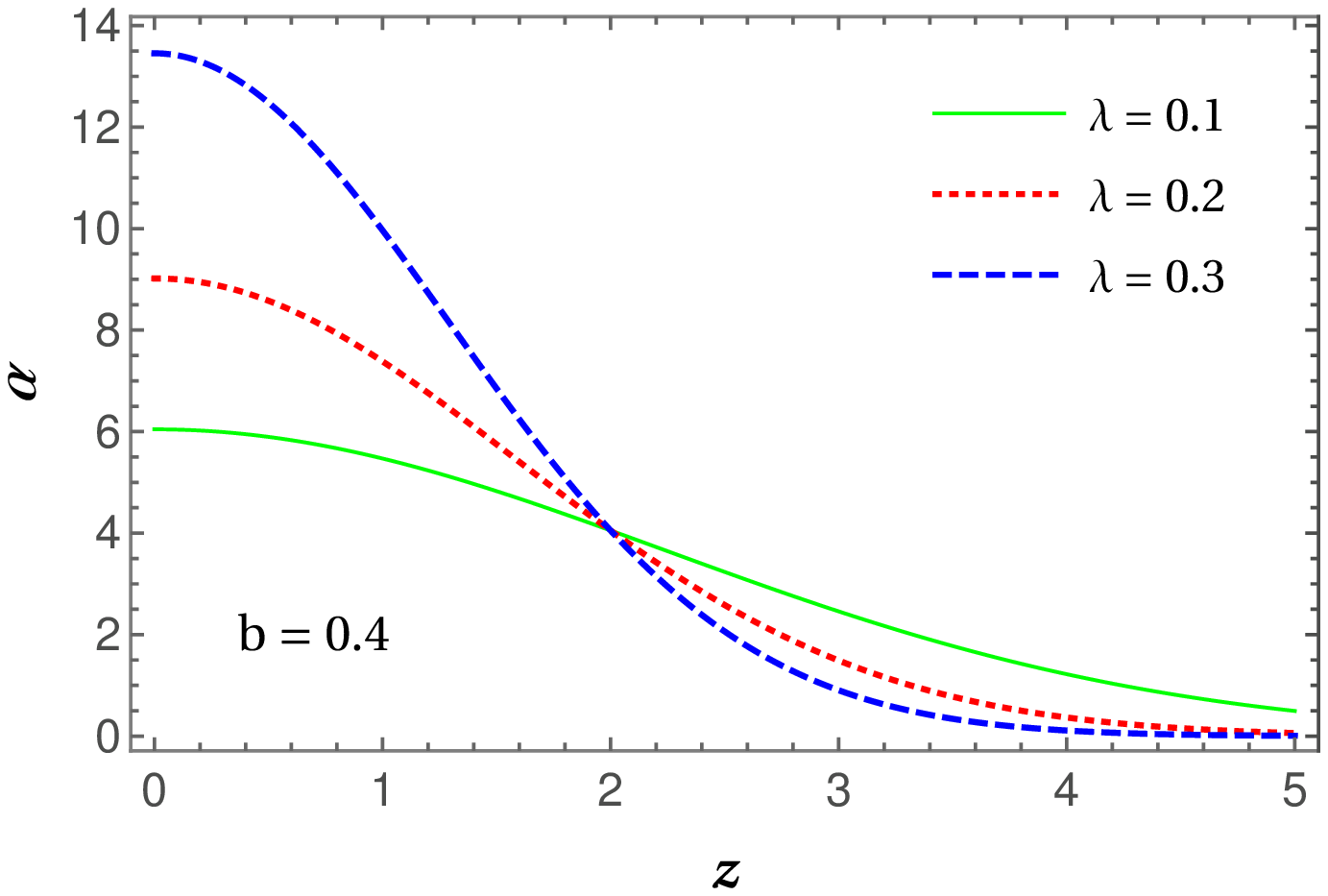}
\end{minipage}
\caption{Considering $\alpha$ versus $z$ coordinate at (a) different temperatures  and (b) different scale parameter.}
\label{fig:Alpha_vs_zfinT}
\end{figure}
\begin{figure}[h!]
\begin{center}$
\begin{array}{cccc}
\includegraphics[width=100 mm]{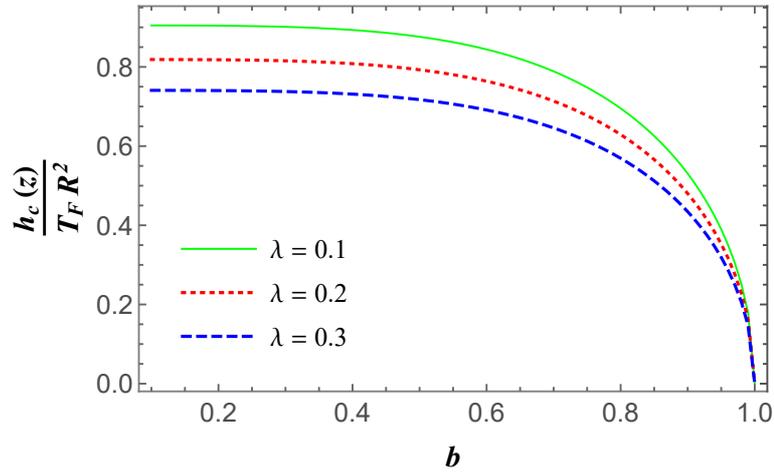}
\end{array}$
\end{center}
\caption{Considering behaviour of critical field versus  temperature.}
\label{fig:critical_vs_b_finiteT}
\end{figure}

As it is represented in  Fig.~\ref{fig:critical_vs_b_finiteT} the critical field has a monotonous manner in low temperature.  On one hand by increasing temperature, the critical field falls down. On the other hand increasing scale parameter decreases the value of critical field. So, at the same temperature greater $\lambda$ leads to smaller critical field which should be obtained for starting point of the pair production. When the probe brane is in near horizon limit,  all the plots with different $\lambda$ coincide and increasing $\lambda$ does not have any effect any more.



According to (\ref{eq:vcp zero}) total potential has been shown in Fig. \ref{fig:x_vs_vtot_finiteT.eps}. We study total potential at fixed value of scale parameter in the Fig. \ref{fig:x_vs_vtot_finiteT.eps} (a) and at fixed temperature in the Fig. \ref{fig:x_vs_vtot_finiteT.eps} (b).
We observe that in Fig. \ref{fig:x_vs_vtot_finiteT.eps} (a) increasing value of $b$ slightly reduces the potential as seen in Table. 2.
and in Fig. \ref{fig:x_vs_vtot_finiteT.eps} (b) increasing value of $\lambda$ has the effect of reducing the potential barrier.
Although increasing both $\lambda$ and $b$ reduces the potential but in comparison the effect of $b$ is not as significant as $\lambda$.
Thus we can say that the effect of temperature is to just strengthen the effect of $\lambda$.

\begin{figure}[h!]
\begin{minipage}[c]{1\textwidth}
\tiny{(a)}\includegraphics[width=7cm,height=6cm,clip]{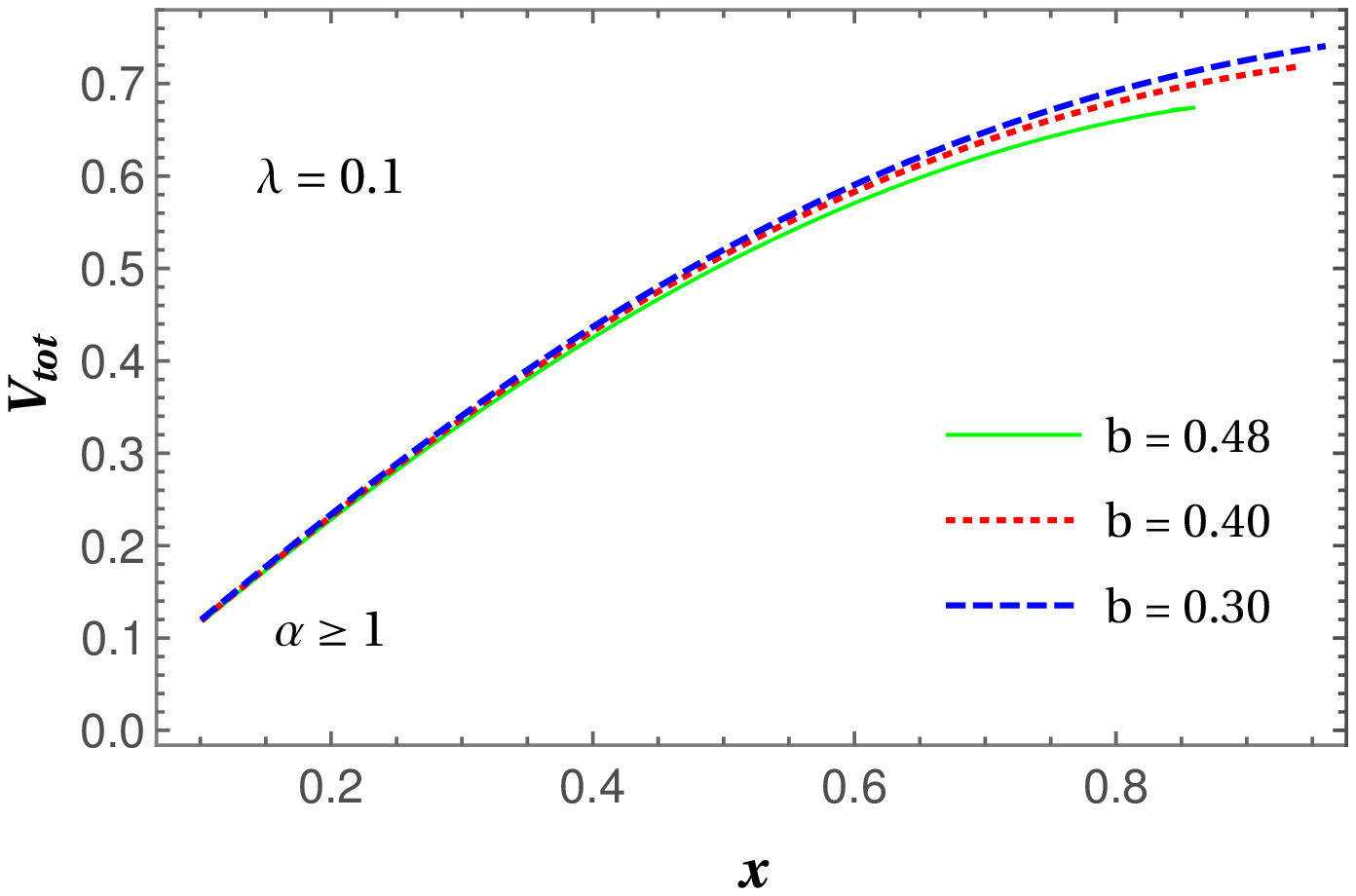}
\hspace{0.1cm}
\tiny{(b)}\includegraphics[width=7cm,height=6cm,clip]{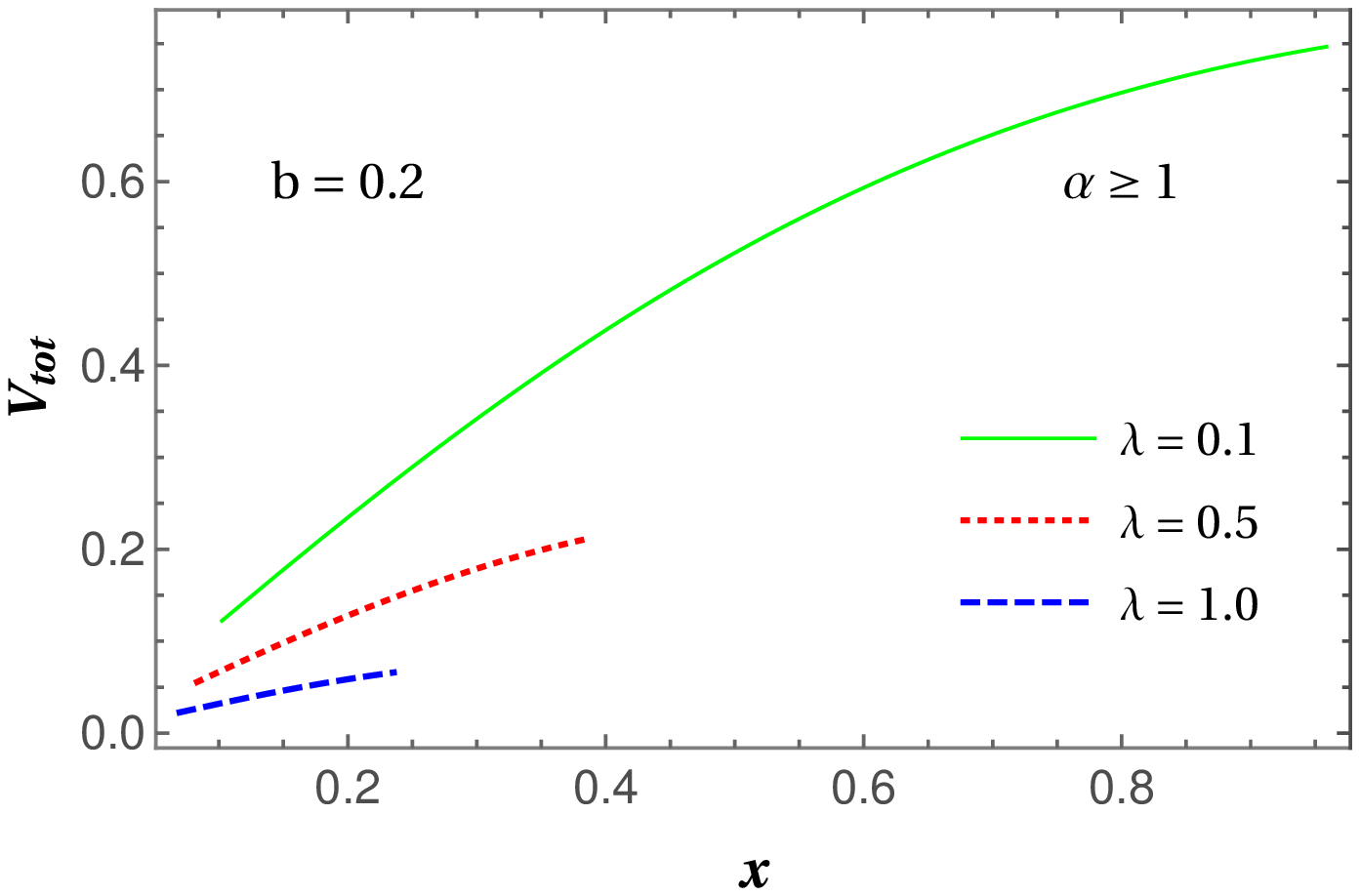}
\end{minipage}
\caption{Considering total potential in pair production process, at (a) different values of temperature and (b) different values of $\lambda$.}
\label{fig:x_vs_vtot_finiteT.eps}
\end{figure}

\begin{table}[ht]
\centering
\begin{tabular}[t]{lccc}
\toprule
~~$b$~~ & $x$~~ & $V$~~ \\
\midrule
~0.2 & 0.38   &  0.21\\
~0.4 & 0.39   & 0.21\\
~0.6 & 0.42   &  0.20\\
~0.8 & 0.26   & 0.14 \\
~0.98 & 0.09   & 0.02 \\
\bottomrule
\end{tabular}
\label{table2}
\caption{Maximum values of  $V_{\mathrm{tot}}$ and $x$ for different values of $b$ for $\lambda=0.5$.}
\end{table}%

\section{Effects of scale parameter on pair production rate at zero temperature } 
\label{section:ppzero}
The  production rate P (per unit time and volume) is evaluated by computing the
expectation value of a circular Wilson loop on the probe brane in the holographic description with the string action \cite{ppr}. According to \cite{coce} we have to find the minimal action, because the pair production probability is given by $\omega\propto e^{-S_{min}}$. In other words, based on \cite{stp},  exponential dependence of the probability rate is given by the minimum of the string effective action.

By the holographic set up, we should consider the action in both zero and finite temperature cases. Deriving differential equation of motion,  we will find numerically the $z(\sigma)$ satisfying the related boundary conditions. Then we will evaluate the action at this specific $z(\sigma)$. So, just to remind the action at zero temperature is defined as,
\begin{equation}\label{SzeroT}
S=2\pi T_F R^2 \int_{0}^{x} d\sigma \frac{1}{z^2}e^{-\lambda z^2}\sqrt{1+z'^2}.
\end{equation}
From the relation (\ref{eq:L zero}) and by  Euler-Lagrange equation,
\begin{equation}\label{eq:Eu-La}
\frac{d}{d \sigma}\left(\frac{\partial  \mathcal{L}}{\partial  z'}\right)-\frac{\partial \mathcal{L}}{\partial z}=0,
\end{equation}
the following differential equation is obtained,
\begin{equation}\label{diffzeroT}
zz''+2(1+z'^2)(1+\lambda z^2)=0,
\end{equation}
where $z=z(\sigma)$ and $z'=\frac{dz(\sigma)}{d\sigma}$.
Now, we should find numerically $z(\sigma)$ satisfying these differential equations and conditions,
\begin{eqnarray}\label{BoundaryzeroT}
z(0)&=&z_{\ast},\nonumber\\
z(\sigma_0)&=&z_0 .
\end{eqnarray}
After finding $z(\sigma)$ the classical action should be evaluated numerically. 

\begin{figure}[h!]
\begin{center}$
\begin{array}{cccc}
\includegraphics[width=100 mm]{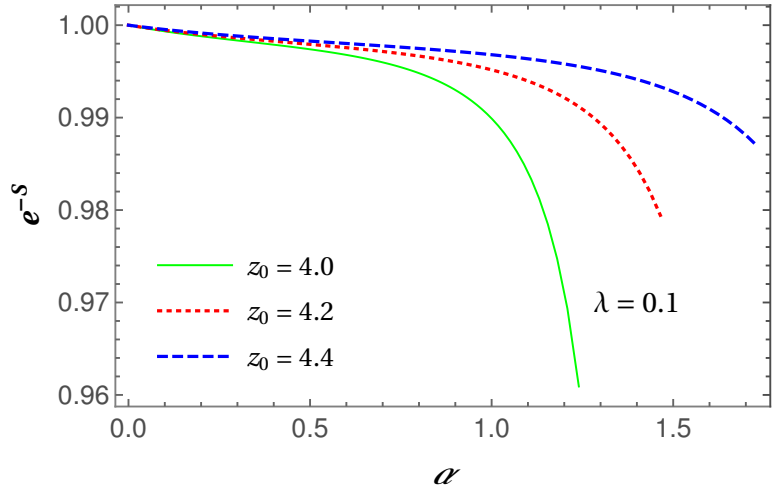}
\end{array}$
\end{center}
\caption{Considering  pair production rate  at zero temperature and fixed value of scale parameter and different probe brane position.}
\label{fig:SandesvsAlpha_fixedLambda_differentz0-alphagreater1.eps}
\end{figure}


 In Fig. \ref{fig:SandesvsAlpha_fixedLambda_differentz0-alphagreater1.eps} pair production rate in the Schwinger effect at zero temperature is represented. 
 Pair production rate is also under effect of probe brane position intensively, as greater value of $z_0$ makes larger pair production rate. However all the plots are definable in a specific width of the $\alpha$ before falling down. When the pair production starts, the rate decreases with increasing $\alpha$ immediately. In other words, although the ratio of the field to its critical value is increasing, but it won't work as an effective  factor of increasing pair production rate, on contrary the pair production rate decreases immediately and in this case there is no catastrophic pair production.  One can interpret that pair production is considerable in special large enough value of $\lambda$. 
Vacuum stability in both limit $\lambda \rightarrow 0$ and $\lambda \rightarrow \infty$ is in agreement  with the stability of vacuum described by $AdS_5$.

\begin{figure}[h!]
\begin{center}$
\begin{array}{cccc}
\includegraphics[width=100 mm]{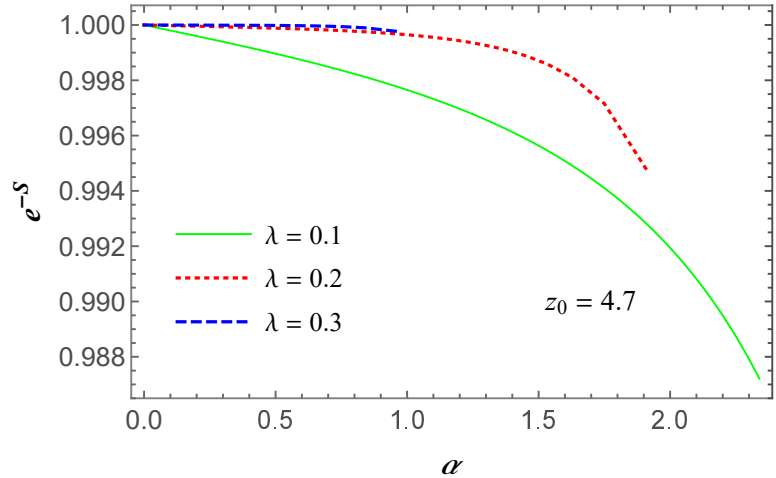}
\end{array}$
\end{center}
\caption{Considering  pair production rate at zero temperature and fixed probe brane position  and different values of large scale parameter.}
\label{fig:SandesvsAlpha_fixedz0_differentlargeLambda.eps}
\end{figure}


In Fig.  \ref{fig:SandesvsAlpha_fixedz0_differentlargeLambda.eps}, pair production rate of the process is represented.
 We can see that maximum values of the rate is obtained when we increase $\lambda$. In large enough value of scale parameter, the potential barrier will be destroyed that is in agreement with our discussion in last section. Studying this plot, we find that from tunneling process  to pair production this is what happens: during tunneling process  there is a maximum value of pair production rate with an approximately monotonous manner with respect to $\alpha$. After vacuum decay, still greater scale parameter leads to larger rate. But this rate falls down after a while more intense than for smaller $\lambda$ cases. Therefore when  $\lambda$ is large enough to destroy potential barrier, pair production via Schwinger effect happens in a short range of $\alpha$ and thereafter its rate fades. The common point in these two plots is that there is no catastrophic pair production  since the creation of the pair has a decreasing behaviour from its maximum value at starting point to zero. So, not this kind of process will continue as long as $\alpha$ is increasing. So we do not consider pair production forever, and no catastrophic pair production happens.
\section{Effects of scale parameter on pair production rate at finite temperature}
\label{sec:pairProduction_fin_Temp}
Considering thermal case from (\ref{eq:L T}) the action is defined as,
\begin{equation}\label{SfinT}
S=2\pi T_F R^2 \int_{0}^{x} d\sigma \frac{1}{z^2}e^{-\lambda z^2}\sqrt{f(z)+z'^2}
\end{equation} 
from Euler-Lagrange equation  the differential equation  is found as, 
\begin{equation}\label{difffinT}
zz'' f(z)-zz'^2 \frac{df}{dz}-\frac{1}{2}zf(z)  \frac{df}{dz}+2(f(z)+z'^2) f(z)(1+\lambda z^2)=0.
\end{equation}
Similar to zero temperature case, classical action at the satisfying $z(\sigma)$ value should be evaluated.

Behaviour of the  pair production rate at finite temperature has been considered in Fig. \ref{fig:SandesvsAlpha_fixedLambdaAndzh_differentz0_finiteT.eps}. Here, fixed value of scale parameter and different temperatures are considered. One can follow  from tunneling process to pair production. The pair production rate decreases with increasing $\alpha$, while greater temperature leads to greater pair production rate at the same scale parameter. In addition, by increasing temperature the pair production rate descends with the smaller slope.  It means that at a fixed $\lambda$ although the pair production rate has a decreasing behaviour similar to zero temperature case, but the temperature can strengthen this rate as larger temperature results in greater pair production rate.
\begin{figure}[h!]
\begin{center}$
\begin{array}{cccc}
\includegraphics[width=100 mm]{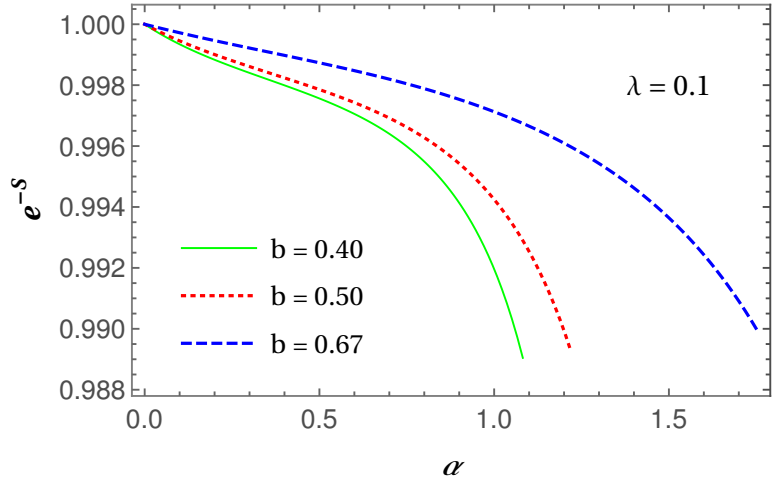}
\end{array}$
\end{center}
\caption{Considering  pair production rate at finite temperature, different probe brane position  and fixed value of  scale parameter.}
\label{fig:SandesvsAlpha_fixedLambdaAndzh_differentz0_finiteT.eps}
\end{figure}

\begin{figure}[h!]
\begin{center}$
\begin{array}{cccc}
\includegraphics[width=100 mm]{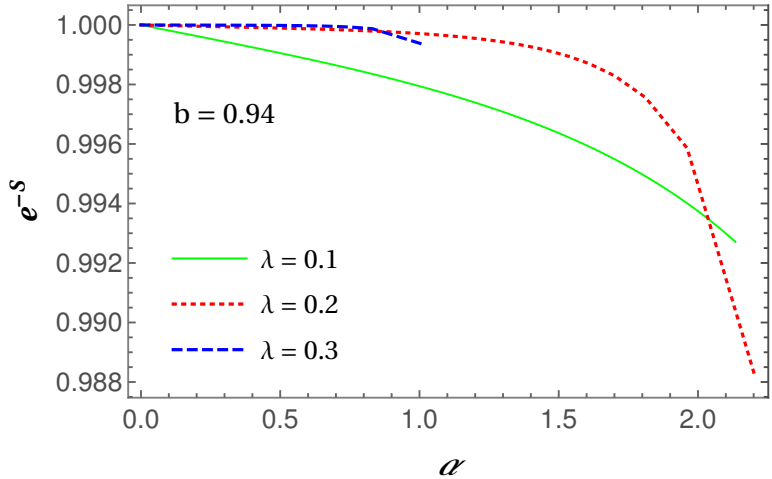}
\end{array}$
\end{center}
\caption{Considering pair production rate at finite temperature and fixed probe brane position  and different large values of scale parameter.}
\label{fig:SandesvsAlpha_fixedz0_differentlargeLambda_finiteT.eps}
\end{figure}


Comparative with Fig. \ref{fig:SandesvsAlpha_fixedz0_differentlargeLambda.eps}  the pair production rate has been represented at different values of scale parameter and finite temperature in Fig. \ref{fig:SandesvsAlpha_fixedz0_differentlargeLambda_finiteT.eps}.
 As we saw in zero temperature case, the maximum value of pair production rate is produced when one manipulates $\lambda$ to increase, as much as possible. In other words, largest scale parameter leads to greater rate of the pair creation. However such a rate is accessible in a limited range of $\alpha$ and it falls down intensively as decreases with increasing $\alpha$. This manner is common in both zero and finite temperature cases. So the temperature does not affect pair production process significantly while probe brane position and scale parameters do that meaningfully. In addition in both zero temperature and finite temperature cases, pair production rate has a decreasing behaviour just after starting the process. Temperature strengthen the pair production rate as we saw in Fig.
\ref{fig:SandesvsAlpha_fixedz0_differentlargeLambda.eps} but it can change the decreasing behaviour and still there is no catastrophic pair production in this Schwinger-like effect.\\
  At the end of this section, one point should be mentioned based on what we have seen so far.
When vacuum decays, the  particle and anti-particle should be created at the same time.  Naturally, they are in short distance from each other, specially in
our case that one of the main result is, system tends to back to stability
immediately. So, according to our previous results
  long distance can not be considered in our
case.

\section{Numerical strategy} 

The integration for $x$ and $V$ respectively are solved numerically to generate the plots shown in Fig.~\ref{fig:x_vs_vtot_z0_1.2_alphage1} and 
Fig.~\ref{fig:x_vs_vtot_finiteT.eps}. 
We choose the value of $z_0 = 1.2$ which falls in the region  $\alpha \ge 1$ as seen in Fig.~\ref{fig:region_Plot_Alpha1}. The upper limit of the integration is varied over a range from $z_{*}^{\mathrm{min}}$ to $z_{*}^{\mathrm{max}}$ to generate the data for the plots.
 The value of $z_{*}^{\mathrm{min}}$ and $z_{*}^{\mathrm{max}}$ used is shown in Table. 3. along with other parameters. The value of $z_{*}^{\mathrm{max}}$ is the value at which $\alpha(\lambda,z_0,z_{*}^{\mathrm{max}} ) = 1$. The integration range for each $\lambda$ value corresponding to $\alpha \ge 1$ is shown in Fig.~\ref{fig:region_Plot_Alpha1} as a black line were the blue  and green dots correspond to  $z_{*}^{\mathrm{min}}$ and  $z_{*}^{\mathrm{max}}$ respectively. We see that the green dots are points on the surface plot were $\alpha = 1$. The region beyond the green dots corresponds to $\alpha < 1$ and hence they set the maximum value for the upper limit in the integration of $x$ and $V$.
We extend the same numerical strategy for the finite temperature plots. The value of the parameters for the finite temperature potential plots are shown in Table. 4.

\begin{figure}[h!]
\begin{center}$
\begin{array}{cccc}
\includegraphics[width=80 mm]{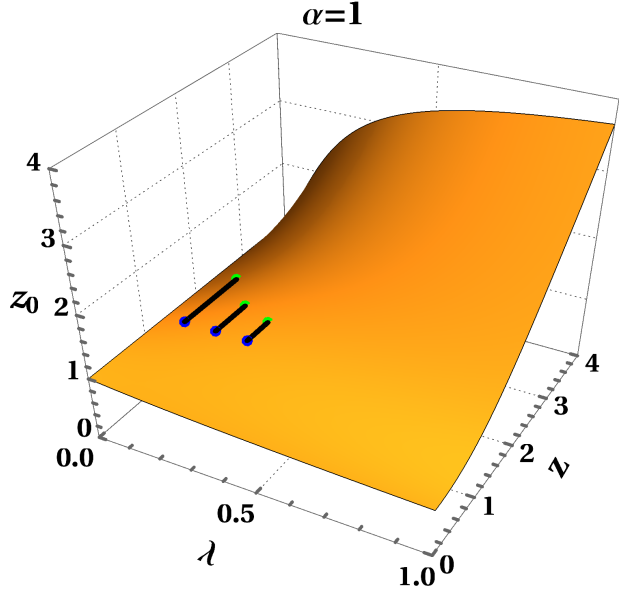}
\end{array}$
\end{center}
\caption{Surface plot satisfying the condition $\alpha = 1$. The region above and below the surface corresponds to $\alpha > 1$ and $\alpha < 1 $ respectively. The black lines show the range of integration for  $\alpha \ge 1$. The blue and green dots correspond to the value of $z_{*}^{\mathrm{min}}$ and  $z_{*}^{\mathrm{max}}$ respectively.}
\label{fig:region_Plot_Alpha1}
\end{figure}

The pair production plots for zero temperature in Sec.~\ref{section:ppzero} are produced by numerically solving the
integration and differential equation shown in Eq.~\ref{SzeroT} and Eq.~\ref{diffzeroT} respectively subject to the boundary condition shown in Eq.~\ref{BoundaryzeroT}. Since now $z(\sigma)$ is a function of $\sigma$, we use an integrated 
($\alpha_{\mathrm{int}}$) value of $\alpha$ given by

\begin{equation}\label{aint}
\alpha_{\mathrm{int}} = \int \limits_0^x d\sigma~\alpha(\lambda,z_0,z(\sigma))
\end{equation}

The differential equation in Eq.~\ref{diffzeroT} is numerically solved over the range of the variable $\sigma$ from 0 to $\sigma_{\mathrm{max}}$. The parameters $z_0$, $\sigma_0$ and $z_{*}$  satisfy the boundary condition defined in Eq.~\ref{BoundaryzeroT}. The data points are generated by varying the upper limit of the integration in Eq.~\ref{SzeroT} and Eq.~\ref{aint} over the range $0$ to $x_{\mathrm{max}}$.
The values of all the parameter used in 
Fig.~\ref{fig:SandesvsAlpha_fixedLambda_differentz0-alphagreater1.eps} and
Fig.~\ref{fig:SandesvsAlpha_fixedz0_differentlargeLambda.eps} is shown in Table.~\ref{table5}.
Again we apply the same strategy for the finite temperature case to numerically solve Eq.~\ref{SfinT} and Eq.~\ref{difffinT}. The parameter values used in Fig.~\ref{fig:SandesvsAlpha_fixedLambdaAndzh_differentz0_finiteT.eps} and Fig.~\ref{fig:SandesvsAlpha_fixedz0_differentlargeLambda_finiteT.eps} are summarized in Table.~\ref{table6}.

\begin{table}[ht]
\centering
\begin{tabular}[t]{lccccc}
\toprule
  ~~$\lambda$~~ & $z_0$ & $z_{*}^{\mathrm{min}}$~~ & ~~$z_{*}^{\mathrm{max}}$  &\\
\midrule
 0.10 &  1.2 &  1.21  &  2.29 &\\
 0.20 &  1.2 &  1.21  &  1.84 &\\
 0.30 &  1.2 &  1.21  &  1.65 &\\
\bottomrule
\end{tabular}
\caption{Parameter values used in  Fig.~\ref{fig:x_vs_vtot_z0_1.2_alphage1}.}
\label{table3}
\end{table}%

\begin{table}[ht]
\centering
\begin{tabular}[t]{lccccccc}
\toprule
 ~~ & $\lambda$~~ & $z_0$ & $z_h$ & $b$ & $z_{*}^{\mathrm{min}}$~~ & ~~$z_{*}^{\mathrm{max}}$  &\\
\midrule
   ~~& 0.10  &  1.2 & 2.5 & 0.48 & 1.21   &  2.31 &\\
Fig.~\ref{fig:x_vs_vtot_finiteT.eps} (a) ~~& 0.10  &  1.2 & 3.0 & 0.40 & 1.21   &  2.28 &\\
   ~~& 0.10  &  1.2 & 4.0 & 0.30 & 1.21   &  2.26 &\\
   ~~& 0.10  &  1.2 & 6 & 0.20 & 1.21   &  2.26 &\\
Fig.~\ref{fig:x_vs_vtot_finiteT.eps} (b) ~~& 0.50  &  1.2 & 6 & 0.20 & 1.21   &  1.47 &\\
   ~~& 1.00  &  1.2 & 6 & 0.20 & 1.21   &  1.34 &\\
\bottomrule
\end{tabular}
\caption{Parameter values for finite temperature potential plots}
\label{table4}
\end{table}%

\begin{table}[ht]
\centering
\begin{tabular}[t]{lccccccc}
\toprule
 ~~~ & $\lambda$~ & $z_0$ & $z_{*}$ & $\sigma_0$ & $\sigma_{\mathrm{max}}$~ & $x_{\mathrm{max}}$  &\\
\midrule
    ~~& 0.10  &  4.0 & 5 & 1.6 & 10  &  1.70 &\\
Fig.~\ref{fig:SandesvsAlpha_fixedLambda_differentz0-alphagreater1.eps}  ~~& 0.10  &  4.2 & 5 & 1.6 & 10  &  1.70 &\\
   ~~& 0.10  &  4.4 & 5 & 1.6 & 10  &  1.70 &\\
\midrule
   ~~& 0.10  &  4.7 & 5 & 0.6 & 10  &  1.00 &\\
Fig.~\ref{fig:SandesvsAlpha_fixedz0_differentlargeLambda.eps}  ~~& 0.20  &  4.7 & 5 & 0.6 & 10  &  0.81 &\\
   ~~& 0.30  &  4.7 & 5 & 0.6 & 10  &  0.68 &\\   
\bottomrule
\end{tabular}
\caption{Parameter values used for plots in Sec.~\ref{section:ppzero}}
\label{table5}
\end{table}%

\begin{table}[ht]
\centering
\begin{tabular}[t]{lccccccccc}
\toprule
 ~~~ & $\lambda$~ & $z_0$ & $z_{*}$ & $z_h$ & $b$ & $\sigma_0$ & $\sigma_{\mathrm{max}}$~ & $x_{\mathrm{max}}$  &\\
\midrule
   ~~& 0.10  &  4.0 & 5 & 10 & 0.40 & 1.7 & 10  &  1.7 &\\
Fig.~\ref{fig:SandesvsAlpha_fixedLambdaAndzh_differentz0_finiteT.eps}  ~~& 0.10  &  4.0 & 5 & 8 & 0.50 & 1.7 & 10  &  1.7 &\\
   ~~& 0.10  &  4.0 & 5 &  6 & 0.67 & 1.7 & 10  &  1.7 &\\
\midrule
   ~~& 0.10  &  4.7 & 5 & 10 & 0.94 & 0.7 & 10  &  1.00 &\\
Fig.~\ref{fig:SandesvsAlpha_fixedz0_differentlargeLambda_finiteT.eps}  ~~& 0.20  &  4.7 & 5 & 10 & 0.94 & 0.7 & 10  &  0.91 &\\
   ~~& 0.30  &  4.7 & 5 & 10 & 0.94 & 0.7 & 10  &  0.77 &\\   
\bottomrule
\end{tabular}
\caption{Parameter values used for plots in Sec.~\ref{sec:pairProduction_fin_Temp}}
\label{table6}
\end{table}%

\newpage
\section{Conclusion}
In this paper we studied condition required for vacuum instability in a holographic theory with dilaton field for both zero and finite temperature cases.
We started by considering a LFH metric background containing dilaton field.
We followed the approach of reference  \cite{uah} while the significant difference between this work and other studies on Schwinger-like effect is that, there is no explicit external field responsible for vacuum decay rather the field is within the metric that causes the vacuum decay.

Potential analysis has been considered at both zero temperature and finite temperature. 
In zero temperature case we have considered that in near boundary region, larger value of critical field should be obtained for pair production, in addition effect of scale parameter is to decrease the value of critical field. Interestingly, in far from the boundary region this scale parameter has no effect. 
The critical field has the maximum value near boundary region and this maximum value decreases with increasing value of the scale parameter. 
Interestingly, in far from the boundary region, the critical field becomes almost independent of the scale parameter.
$\alpha$  is the ratio of dilaton field to its critical value and we have found that there is  a preferable  region  where the condition $\alpha\geq 1$ is obtainable  almost irrespective of $\lambda$ value. Also, $\alpha$ depends on probe brane position significantly. 

During tunneling process at zero temperature, increasing scale parameter leads to diminishing potential barrier, so although the pair production via Schwinger effect has not been started yet, but pair creation according to tunneling process increases. In current study, for pair creation via Schwinger effect, condition $\alpha\geq 1$ is not enough because we need $\lambda$ to be large. Similar to zero temperature case, at finite temperature also $\alpha$ has its maximum value in near boundary region. In the region far from the boundary, $\alpha$ becomes independent of the temperature. The effect of scale parameter on $\alpha$ depends on the region completely, as the effect of scale parameter near the boundary is completely different with near horizon region. 
We found that the low temperature region does not have significant affect on the critical field, but at high temperature the critical field falls down rapidly and with further increases in temperature it become independent of the scale parameter. 
The results for the  potential analysis at finite temperature is similar with the zero temperature case.

Pair production rate has been considered as the exponential function of the action. We observe that in both zero temperature and finite temperature cases, the pair production rate decreases with increasing values of scale parameter and probe brane position. Also the rate of decrease is rapid beyond the region $\alpha \ge 1$. So, in this work although we observe vacuum instability under very restricted condition but there is no catastrophic pair production.

\textbf{Acknowledgement}\\
This work was supported by the National Natural Science Foundation of
China (Grant No. 11575254), the National Key Research and Development Program of China (No. 2016YFE0130800).  
SN is supported by the China Postdoctoral Council under the International Postdoctoral Exchange Fellowship Program. 
\\

\end{document}